\documentclass[journal]{IEEEtran}
\usepackage{xcolor}
\usepackage{cite}
\usepackage{amsmath}
\usepackage{algorithm}
\usepackage{algorithmic}
\usepackage{multirow}
\usepackage{graphicx}
\usepackage{textcomp}
\usepackage{amssymb}

\usepackage{lineno,hyperref,diagbox,multicol,balance,epstopdf}
\ifCLASSINFOpdf

\else

\fi

\begin{document}
	
	\title{Disentangling Style and Speaker Attributes for \\TTS Style Transfer}

	\author{Xiaochun An, ~\IEEEmembership{IEEE Student Member} \\
        Frank K. Soong, ~\IEEEmembership{IEEE Fellow} \\
		Lei Xie$^\star$, ~\IEEEmembership{IEEE Senior Member}
		\thanks{Xiaochun An and Lei Xie are with the Audio, Speech and Langauge Processing Group (ASLP@NPU), School of Computer Science, Northwestern Polytechnical University, Xi'an, China (email: xiaochunan@npu-aslp.org, lxie@nwpu.edu.cn)}
		\thanks{Frank K. Soong is with the Speech Group, Microsoft Research Asia (MSRA), Beijing, China (email: frankkps@microsoft.com)}
		\thanks{$^\star$ Corresponding author}
	}

	% The paper headers
	\markboth{Journal of \LaTeX\ Class Files,~Vol.~, No.~, January~2022}
	{Shell \MakeLowercase{\textit{et al.}}: Bare Demo of IEEEtran.cls for IEEE Journals}
	
	% make the title area
	\maketitle
	\begin{abstract}
      End-to-end neural TTS has shown improved performance in speech style transfer. However, the improvement is still limited by the available training data in both target styles and speakers. Additionally, degenerated performance is observed when the trained TTS tries to transfer the speech to a target style from a new speaker with an unknown, arbitrary style. In this paper, we propose a new approach to seen and unseen style transfer training on disjoint, multi-style datasets, i.\,e., datasets of different styles are recorded, one individual style by one speaker in multiple utterances. An inverse autoregressive flow (IAF) technique is first introduced to improve the variational inference for learning an expressive style representation. A speaker encoder network is then developed for learning a discriminative speaker embedding, which is jointly trained with the rest neural TTS modules. The proposed approach of seen and unseen style transfer is effectively trained with six specifically-designed objectives: reconstruction loss, adversarial loss, style distortion loss, cycle consistency loss, style classification loss, and speaker classification loss. Experiments demonstrate, both objectively and subjectively, the effectiveness of the proposed approach for seen and unseen style transfer tasks. The performance of our approach is superior to and more robust than those of four other reference systems of prior art.
	\end{abstract}
	
	
	\IEEEpeerreviewmaketitle

	\section{Introduction}
	\label{Introduction}
	\IEEEPARstart{E}{nd-to-end} neural TTS models, such as Tacotron 2~\cite{shen2018natural}, can produce high quality speech with naturalness close to that of human speakers~\cite{wang2017tacotron, ping2018deep, fengyu2019improving, guangzhi2020generating}. The neural TTS models usually consist of an encoder-decoder neural network~\cite{sutskever2014sequence, bahdanau2015neural} which is trained to map a given text sequence to a sequence of speech frames. Extensions of these models have shown that the \textit{speech styles} (e.g., speaker identity, speaking style, emotion and prosody) can be modelled and controlled in an effective way~\cite{wu2019end-to-end, stanton2018predicting, liu2020expressive, lei2021fine}. Many TTS application scenarios, such as audiobook narration, news broadcasting, and conversational assistants, demand single-speaker, multi-style speech synthesis, i.\,e., a TTS to speak simultaneously in multiple styles. However, the corresponding performance in this area is inadequate due to the lack of single-speaker, multi-style speech data in general.
	
	Currently most neural TTS models~\cite{akuzawa2018expressive, xiaochun2019learning, hsu2018hierarchical, habib2020semi-supervised, guangzhi2020fully} are trained with an expressive, single-style corpus. Acquisition of a large set of single-speaker speech data with multiple styles, which is useful for training a good neural expressive TTS, is usually expensive and time consuming. Alternatively, a more effective solution is to perform \textit{speech style transfer}~\cite{pan21d_interspeech}, which allows a speaker to learn the desired style from the speech data in the same style but recorded by other speakers and preserves the target speaker's timbre. The neural TTS models with a reference encoder~\cite{skerry2018towards, li2021controllable}, global style tokens (GST)~\cite{wang2018style}, and a variational autoencoder (VAE)~\cite{zhang2019learning} have become popular for controlling and transferring \textit{speech styles}. Theoretically, these models can model any complex styles in a continuous latent space, hence one can control and transfer style by manipulating the latent variables or variational inference from a reference audio. However, these approaches tend to have too much entangled information to make the style rendering robust and interpretable, and independent control of specific speech characteristics (e.g., speaker identity and speaking style) is not clear and direct. In style transfer, one needs to transfer all styles whether desired or not, which may not fit the contexts thus hurts generalization. When conducting style control, one can hardly find the direct relationships between the target styles and the parameters in the embedding vector of the style representations to facilitate a direct control strategy.

To address the above issues, Bian \emph{et al.}~\cite{bian2019multi} introduce a multi-reference encoder to GST~\cite{wang2018style} to model multiple styles simultaneously and adopts intercross training to extract and separate different classes of speech styles. Occasionally, the model shows a successful style transfer, but the intercross training does not guarantee each possible combination of style classes is seen during training, leading to a missed opportunity to learn disentangled representations of styles and sub-optimal results on disjoint, multi-style datasets. In ~\cite{whitehill2020multi}, the authors address the challenges of multi-reference style transfer on disjoint datasets by using an adversarial cycle consistency training scheme. Different from intercross training, their training tries all possible combinations of style classes via paired and unpaired triplets. Thus results in the disentanglement of multiple style dimensions and classes, and enables the style transfer to be more faithful than other existing methods.

Although the style transfer performance has been improved in ~\cite{whitehill2020multi}, it is still limited to a style transfer from a speaker seen in training, but inadequate to transfer to a target style from a new speaker with an unknown, arbitrary style. In addition, collecting training samples needed of new styles is always challenging and labor-intensive. Transferring style from one dataset to another (i.\,e., disjoint, multi-style datasets) is an appealing feature for a TTS system. Unseen style transfer on disjoint, multi-style datasets needs to be further improved.
	
	In this paper, we propose an encoder-decoder neural network to improve \textit{performance of seen and unseen style transfer} on disjoint, multi-style datasets. Our preliminary work has been presented in ~\cite{An2021improving}, in which how to distinguish different style types and capture the characteristics of individual speakers are not explicitly considered. In the current work, our preliminary work is extended with explicit constraints of different style types and different speakers to learn more discriminative style and speaker representations, which demonstrates improved discrimination of different styles and speakers. The modules for learning the style and speaker representations are sharpened in the current work, and the improvement has been confirmed in the Section~\ref{Comparison of style transfer without style/speaker classifier} and Section~\ref{Effect of different objectives on style transfer}.
	
	\textit{Our first contribution is to adopt an inverse autoregressive flow (IAF) to improve variational inference and learn discriminative style representations.} Previous style embeddings~\cite{hsu2018hierarchical, zhang2019learning} are obtained by adopting the VAE~\cite{kingma2014auto-encoding} network. On the basis of the mean-field approximation, VAE assumes the independence of utterances and models them with a corresponding isotropic, Gaussian latent space. Despite the tractability of computation, the dimension-independent Gaussian distribution is not expressive enough, which has been investigated in various work~\cite{atanov2018deep}, and utterances with the same style are not dependent but connected by sharing one global style space. Hence, we introduce an IAF in our proposed network to perform the variational inference. In the IAF model, posterior distributions are formulated by a series of cascaded invertible transformations to map a simple initial density to an arbitrarily complex, flexible distribution with tractable Jocabians~\cite{kingma2016improving}. Thus a flexible approximate distribution can be used for discriminative style embeddings. Note that IAF has been recently applied to speech processing tasks, e.g., fast and high-fidelity WaveNet based speech synthesis, Oord \emph{et al.}~\cite{den2018parallel} propose to use probability density distillation as a bridge between trained WaveNet (teacher model) and IAF (student model). Esling \emph{et al.}~\cite{esling2019universal} propose a universal audio synthesizer built with normalizing flows~\cite{Aggarwal2020using} to learn the latent space representation for semantic control of a synthesizer by interpolating latent variables. In this study, IAF is used as normalizing flows to perform the variational inference for learning discriminative and expressive style embeddings.
		
	\textit{Our second contribution is to develop speaker encoder network for joint training of our proposed network to learn discriminative speaker representations.} In recent years, researchers usually adopt a speaker recognition model~\cite{snyder2018x} or a speaker verification model~\cite{wan2018generalized} to learn the speaker representations. Although they can extract the speaker embeddings, such speaker extractor models always need to be pretrained by using an independent dataset in advance. If the training data is inadequate, as shown in ~\cite{Ye2018Transfer}, it leads to poor speaker representations. Therefore, we introduce a well-designed speaker encoder in our proposed network to train jointly with the rest network to learn discriminative speaker representations from any speakers, even if the speaker are not seen in the training data.
	
       \textit{Our third contribution is to use six specifically-designed losses in network training.} The style transfer accuracy and speaker preservation are both considered in our proposed approach of seen and unseen speech style transfer. The network is trained to learn more discriminative style and speaker representations in a disentangled manner, by optimizing the tradeoff between style transfer accuracy and speaker identity preservation through the six specifically-designed objectives: reconstruction loss, adversarial loss, style distortion loss, cycle consistency loss, style classification loss, and speaker classification loss. The reconstruction loss is used to measure the distortions in both source and target reconstructions; the adversarial loss to ``fool" a well-trained discriminator; the style distortion loss to constrain the style representation of a source utterance to be closer to the target style representation; the cycle consistency loss to ensure that the transferred utterance can preserve the speaker identity of the source utterance; the style classification loss and the speaker classification loss to further improve their style and speaker representations and to make the category of each style and each speaker more distinguishable. In this way, we can transfer the speech to a target style from a new speaker with an unknown, arbitrary style, which does not need to be seen in training.
	
	Subjective and objective experiments show that our approach to seen and unseen speech style transfer can improve 1) speech naturalness, 2) style similarity, and 3) speaker similarity, as compared with the four reference systems of prior art. Specially, on the unseen style transfer task, the reference systems, in most cases, fail to transfer an unseen style to a target style, and are not effective in preserving the speaker's timbre, resulting in lower similarity scores.
	
	To summarize our contributions, this paper proposes a novel approach to seen and unseen speech style transfer that can significantly improve performance of seen and unseen style transfer on disjoint, multi-style datasets. Specifically, the network is trained to minimize six specifically-designed losses to ensure the style representation of a source utterance is closer to the target style representation after the transfer and the transferred utterance can preserve the speaker identity of the source utterance, even when an utterance to be transferred is from a new speaker with an arbitrary, unknown style. In addition, the proposed scheme can be used as a data augmentation method to generate a single-speaker, multi-style speech data, which is useful for various speech applications, such as multi-style TTS and voice conversion.

	The rest of this paper is organized as follows. Section~\ref{Related works} reviews the previous studies on single-reference and multi-reference speech style transfer. Section~\ref{Seen and unseen style transfer} presents our proposed approach to seen and unseen speech style transfer on disjoint, multi-style datasets. Section~\ref{ExperimentalSetup} and Section~\ref{ExperimentalResults} introduce the experimental setup and results, respectively. Section~\ref{Conclusion} concludes this paper and discusses our future work.

	\section{Related works}
	\label{Related works}
    Currently, there are mainly two major approaches, supervised and unsupervised to speech style transfer in TTS. The supervised approach, which takes manual style labels as additional TTS model input, has been shown effective in multi-speaker TTS~\cite{gibiansky2017deep}. However, this approach can hardly deal with more complex styles like different speaking styles, varying emotion levels and prosodic changes, etc., because there are no clean objective measures to annotate these styles. The unsupervised approach, which combines neural TTS models and a single-reference or multi-reference encoder, is more popular than the supervised approach. In this section, we will briefly review the related work on single-reference and multi-reference speech style transfer.
	\subsection{Single-reference Speech Style Transfer}
	\label{Single-reference speech style transfer}	
	Skerry-Ryan \emph{et al.} extend the Tacotron architecture~\cite{wang2017tacotron} by adding a reference encoder module~\cite{skerry2018towards, gururani2019prosody} that compresses the style of a variable-length audio signal into a fixed-length vector, as the reference embedding. Here, the reference encoder is to learn the embedding space of style from the speech data directly in the training process. The learned embedding, when used as a condition in synthesis, can generate speech signals with a style similar to that of the reference signal, even when the reference and target speakers are different.

Another popular single-reference speech style transfer is the GST model~\cite{wang2018style}, which augments the reference encoder by utilizing a multi-head attention~\cite{vaswani2017attention} based style token layer to extract rich style information in the training data. The extracted information is then used to control the synthesis, such as varying speed and speaking style. Similarly, it can be used for style transfer, replicating the speaking style of a single audio clip across a long-form text corpus.

Deep generative models, such as VAE~\cite{kingma2014auto-encoding} and GAN~\cite{Ian2014generative, Han2018Self-Attention}, are powerful architectures which can learn complicated distributions in an unsupervised manner. Particularly, VAE, which can explicitly model the corresponding latent variables, has become one of the most popular schemes and achieved significant advancement in text generation~\cite{bowman2015generating}, image generation~\cite{higgins2016beta, burgess2018understanding} and speech generation~\cite{akuzawa2018expressive, hsu2017learning}. Zhang \emph{et al.}~\cite{zhang2019learning} introduce the VAE in the neural TTS model, to learn the latent representations of speaking style in an unsupervised manner. The learned style representations are then fed into a TTS network to control the style of the synthesized speech.

Theoretically, the above approaches can model any complex styles in a continuous latent space, so that one can control the transferred style by manipulating the latent variables or conducting a variational inference from a reference audio. However, these methods model all speech styles into one single representation, which is not versatile enough to control specific speech attributes independently. When conducting style transfer, one then has to transfer all the embedded styles, desired or not, which may not fit well with the contexts and may hurt its generalization. In addition, these approaches fail to generalize to a new domain which is unseen in training. For example, to create speech in different speaker identities and speaking style classes by using a single model, a dataset containing audio samples for each speaking style class and speaker identity is needed, and yet the model can still fail to transfer the speaking style from a new speaker with an arbitrary, unknown style.

	\subsection{Multi-reference Speech Style Transfer}
	\label{Multi-reference speech style transfer}
	The single-reference speech style transfer has primarily focused on the transfer of a single-style reference audio sample. Those methods are inadequate for disjoint, multi-style datasets because of their lack of domain adaptation~\cite{daume2006domain} capability.

Recently, Bian \emph{et al.}~\cite{bian2019multi} introduce a multi-reference encoder to GST~\cite{wang2018style} and adopt an intercross training scheme, to ensure that each sub-encoder of the multi-reference encoder disentangles and controls a specific style independently. The model shows successful style transfer in a multi-style scenario. However, its intercross training scheme does not guarantee each combination of style classes is seen in training, leading to a missed opportunity to learn disentangled representations of styles and sub-optimal results on disjoint datasets. To address the above problems, Whitehill \emph{et al.}~\cite{whitehill2020multi} propose an adversarial cycle consistency training scheme to ensure the use of information from all style classes. Different from intercross training, the scheme sweeps across all combinations of style classes via paired and unpaired triplets. This provides disentanglement of multiple style classes, enabling the model to transfer style in a more faithful manner than the existing methods.
	
However, similar to ~\cite{bian2019multi}, the method of ~\cite{whitehill2020multi} still suffers a limitation that can only transfer the style seen in training, and is inadequate to transfer the speech to a target style from a new speaker with an unknown, arbitrary style, thus narrowing down applicable scenarios for neural TTS. In addition, recording training samples in a new style (e.g., poetry style) is challenging and labor-intensive, transferring style from one dataset to another (i.\,e., disjoint, multi-style datasets) is appealing for TTS systems.

More recently, the first author of this paper proposes an approach to perform seen and unseen style transfer on disjoint datasets~\cite{An2021improving}. Despite the realization of seen and unseen style transfer, the method cannot distinguish different style and speaker types well, especially for unseen style transfer, resulting in sub-optional style transfer accuracy and speaker preservation. Hence, it is necessary to further enhance the performance of seen and unseen speech style transfer on disjoint, multi-style datasets.
	
	\section{Proposed approach to seen and unseen speech style transfer}
	\label{Seen and unseen style transfer}	
	
	\subsection{System Overview}
	\label{System overview}
\begin{figure}[!t]	
		\begin{minipage}[b]{1.0\linewidth}
			\centerline{\includegraphics[width=1.0\textwidth]{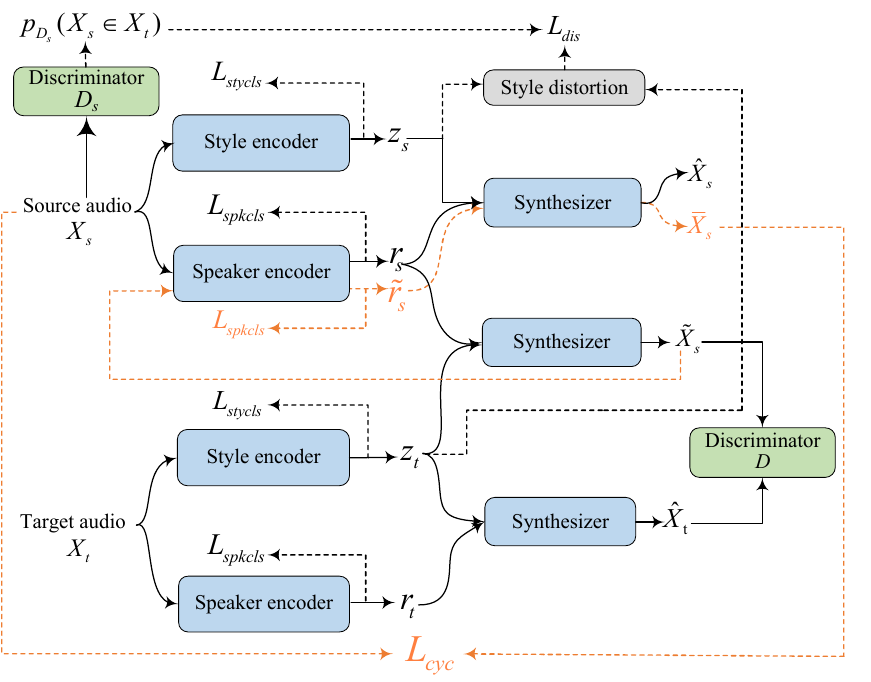}}
		\end{minipage}
		\caption{The diagram of performing seen and unseen speech style transfer.}
		\label{fig1}
	\end{figure}

	Our approach to seen and unseen speech style transfer is built upon an encoder-decoder neural network~\cite{sutskever2014sequence, bahdanau2015neural}. Fig.~\ref{fig1} illustrates our proposed framework for both seen and unseen styles on disjoint, multi-style datasets. Let $X_s =\{x_s^{(1)} ,\ldots, x_s^{(n)}\}$ be the source utterances, and $X_t=\{x_t^{(1)} ,\ldots, x_t^{(m)}\}$, the target utterances, respectively. We assume that each speech utterance $x$ can be decomposed into style representation, $z \in Z$, and speaker representation, $r \in R$. Each source utterance, $x_s^{(i)} \in X_s$, has its individual style, $z_s^{(i)}$, and the target utterance, $x_t^{(j)} \in X_t$, has the style, $z_t^{(j)}$. We use style encoder and speaker encoder, $E_z(x)$ and $E_r(x)$, to learn discriminative style representation $z$ and speaker representation $r$ of an utterance $x$, respectively: $z_s^{(i)}=E_z(x_s^{(i)})$, $r_s^{(i)}=E_r(x_s^{(i)})$, $z_t^{(j)}=E_z(x_t^{(j)})$, $r_t^{(j)}=E_r(x_t^{(j)})$. In this paper, we use Tacotron 2~\cite{shen2018natural} as the synthesizer $T$, which converts the combined encoder states (including style representations, speaker representations, and text encoder states) to generate a target Mel spectrogram with the target style and the target speaker's timbre. The Tacotron 2 is composed of a text encoder and a decoder with attention, whose details will be described in Section~\ref{decoder}. We then employ the LPCNet neural vocoder~\cite{valin2019real, Valin2019LPCNET} to reconstruct the final speech waveforms from the generated Mel spectrograms, and the details of each individual component in LPCNet vocoder will be described in Section~\ref{Network}. Our network is trained with six objective loss functions, including: reconstruction ($\mathcal{L}_{rec}$), adversarial ($\mathcal{L}_{adv}$), style distortion ($\mathcal{L}_{dis}$), cycle consistency ($\mathcal{L}_{cyc}$), style classification ($\mathcal{L}_{stycls}$) and speaker classification ($\mathcal{L}_{spkcls}$) losses. Next, we introduce the learning of latent style space, speaker encoder network, synthesizer network and six objective losses.

	\subsection{Learning Latent Style Space}
	\label{style encoder}%

	\begin{figure}[!t]	
		\begin{minipage}[b]{1.0\linewidth}
			\centerline{\includegraphics[width=1.0\textwidth]{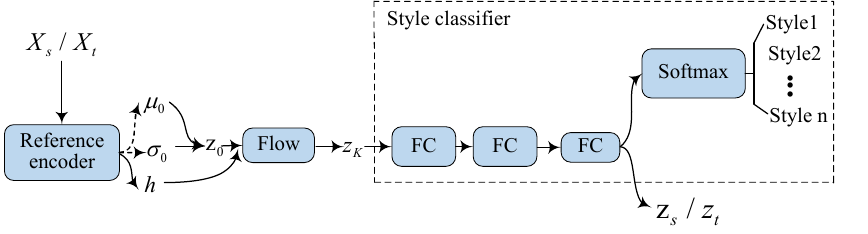}}
		\end{minipage}
		\caption{Architecture of style encoder for learning style representations.}
		\label{fig2}
	\end{figure}

	The style encoder $E_z(x)$ constructs a latent style space, and then outputs a sampled style representation $z$ to the synthesizer $T$ to guide style generation. As discussed above, the generated style space via VAE network~\cite{kingma2014auto-encoding}, usually a Gaussian distribution, may not be expressive enough to transfer the style effectively. In this paper, we resort to the IAF~\cite{kingma2016improving, rezende2015variational}, a potent technique to construct sophisticated distributions, to learn discriminative and expressive style representations. In other words, IAF can map a simple initial variable to a more complex one by applying a chain of cascaded invertible transformations. As shown in Fig.~\ref{fig2}, we let a reference encoder network output $\mu_0$ and $\sigma_0$, in addition to an extra output $h$, which serves as an additional input to each subsequent step in the flow. We draw a random sample $\epsilon\sim\mathcal{N}(0, I)$, and initialize the chain with:
\begin{equation}
	z_0=\mu_0 + \sigma_0 \odot\epsilon
	\label{eq1}
	\end{equation}
Afterward, the initial variable $z_0$ along with hidden output $h$ is provided to $k$ steps of inverse autoregressive transformations to obtain flexible posterior probability distribution with latent variable $z_k$:
\begin{equation}
	z_k=\mu_k + \sigma_k \odot z_{k-1}
	\label{eq2}
	\end{equation}

	\begin{figure}[!t]	
		\begin{minipage}[b]{1.0\linewidth}
			\centerline{\includegraphics[width=0.85\textwidth]{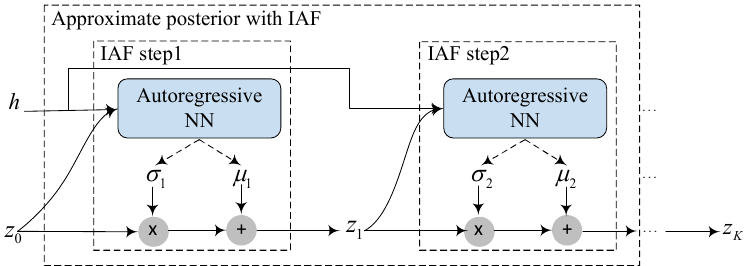}}
		\end{minipage}
		\caption{The process of flow transformations.}
		\label{fig3}
	\end{figure}

In each step of the flow transformations, we adopt an autoregressive neural network with inputs $z_{k-1}$ and $h$, and outputs $\mu_k$ and $\sigma_k$. And amortization is performed by using $h$ as input to autoregressive networks of flow transformations~\cite{berg2018sylvester}. These autoregressive transformations are invertible if $\sigma_i\textgreater0$ condition is satisfied for $i^{th}$ value of $D$ dimension. Autoregressive structure of flow allows simple computation of the Jacobian determinant of each transformation as a change in global posterior probability density of reference encoder network denoted as $\log q(z_K|x)$, where $z_K$ is output of the last flow step. Eq.(\ref{eq3}) provides a tractable change of the probability density, and its detailed derivation can be found in ~\cite{kingma2016improving}. The flexibility of the distribution of the final iteration $z_K$ , and its ability to closely fit to the true posterior, increases with the expressivity of the autoregressive models and the depth of the chain. See Fig.~\ref{fig3} for an illustration of flow transformations.
\begin{equation}
	\log q(z_K|x)=-\sum_{i=1}^{D}(\frac{1}{2}\epsilon_i^2+\frac{1}{2}\log (2\pi)+\sum_{k=0}^{K}\log (\sigma_{k,i}))
	\label{eq3}
	\end{equation}

Different from~\cite{An2021improving}, we then plug a style classifier to the style encoder shown in Fig.~\ref{fig2}, which helps to learn more discriminative style embeddings for better discriminations of different styles. In detail, the classifier includes three fully connected (FC) layers, all with ReLU activation, and a softmax layer to output the probability of different style types, i.\,e., reading, broadcasting, talking, story-telling, customer-service, poetry, and game styles. We use the output of the third FC layer as the style representation, $z_s$ or $z_t$. In this way, we can obtain more discriminative and expressive style embeddings to guide style generation.
	
	\subsection{Developing Speaker Encoder Network}
	\label{speaker encoder}
	In previous works, speaker embeddings are usually obtained by using a speaker recognition model~\cite{snyder2018x, villalba2019state} trained on the voxceleb corpus available in the Kaldi toolkit~\cite{chung2018voxceleb2, povey2011kaldi} or a text-independent speaker verification task~\cite{wan2018generalized, li2018deep}, where the speaker extraction models need to be pre-trained. In our proposed network, we add a well-designed speaker encoder, $E_r(x)$, to learn the speaker representations. Different from ~\cite{Ye2018Transfer}, we propose to jointly train the speaker encoder network $E_r(x)$ with the rest of the neural TTS modules. We expect that the use of the speaker encoder can learn discriminative speaker embeddings to capture the characteristics of individual speakers, even the speaker is not seen in the training data.

\begin{figure}[!t]	
		\begin{minipage}[b]{1.0\linewidth}
			\centerline{\includegraphics[width=1.0\textwidth]{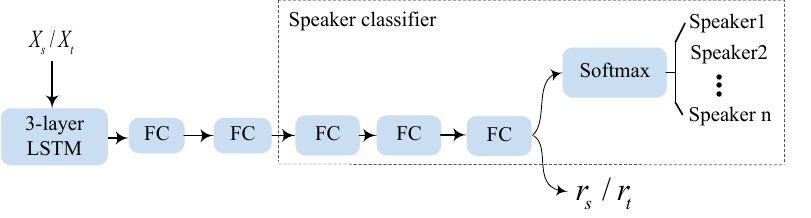}}
		\end{minipage}
		\caption{Architecture of speaker encoder for generating speaker embeddings.}
		\label{fig4}
	\end{figure}

Fig.~\ref{fig4} shows the architecture of speaker encoder in our approach, which maps a sequence of acoustic features of a speech utterance, into a fixed-dimensional embedding vector. First, we feed the features extracted from an utterance $x$ into a 3-layer LSTM network. Two FC layers both with ReLU activation are connected to the last LSTM layer as an additional transformation of the last frame response of the network. Different from~\cite{An2021improving}, a speaker classifier is added into speaker encoder to learn more discriminative speaker representations. As shown in Fig.~\ref{fig4}, the classifier has the same structure as the style classifier, consisting of three FC layers, all with ReLU activation, and a softmax layer to output the probability of different speaker. Similarly, the output of the third FC layer is used as the speaker representation $r_s$ or $r_t$, which is then used as a condition for the synthesizer $T$ to guide speaker identity generation.

\subsection{Synthesizer}
	\label{decoder}	
	In this paper, we leverage Tacotron 2 as the synthesizer $T$. Tacotron 2 is an attention-based sequence-to-sequence network~\cite{sutskever2014sequence, bahdanau2015neural, vinyals2015grammar}, composed of a text encoder and a decoder with attention, which generates a Mel spectrogram as a function of an input text sequence and conditions the signal generated by the auxiliary encoder networks (e.g., style encoder and speaker encoder). It closely follows the network architecture of Tacotron 2~\cite{shen2018natural}. Input phonemes are represented using a learned 512-dim phoneme embedding, which is passes through a stack of three convolutional layers each containing 512 filters with shape $5\times1$, followed by a bidirectional LSTM of 256 units for each direction. In this work, to condition the output on additional attribute representations (e.g., style representations and speaker representations), the resulting text encodings are concatenated with the generated style embeddings $z$ from the style encoder and the extracted speaker embeddings $r$ from the speaker encoder, and then are accessed by the decoder through a location sensitive attention mechanism~\cite{chorowski2015attention-based}, which takes attention history into account when computing a normalized weight vector for aggregation.

The base Tacotron 2 autoregressive decoder network takes the attention-aggregated text encoding, and the bottlenecked previous frame (processed by a pre-net comprised of two FC layers of 256 units) at each step as input. The decoder input is then passed through a stack of two uni-directional LSTM layers with 1024 units. The output from the stacked LSTM is concatenated with the new decoder input (as a residual connection~\cite{wu2016google}), and linearly projected to predict the Mel spectrogram of the current frame, as well as the end-of-sentence token. Finally, the predicted spectrogram frames are passed to a post-net, which predicts a residual that is added to the initial decoded sequence of spectrogram frames, to predict the spectrograms by minimizing the overall mean squared errors.

	\subsection{Six Specifically-designed Objectives}
	\label{Six specifically-designed objectives}	
	For the synthesizer $T$, we form a reconstruction loss $\mathcal{L}_{rec}$ to encourage the utterance from $T$, given style representation $z$ and speaker representation $r$ of an utterance $x$, to reconstruct $x$ itself:
\begin{equation}
%\vspace{-4pt}
\begin{split}
   &\mathcal{L}_{rec}(\theta_{E_z},\theta_{E_r},\theta_{T}) \\
   &=\mathbb{E}_{x_s \sim X_s}[-\log p_{T}(x_s |z_s,r_s)] \\
   &+\mathbb{E}_{x_t \sim X_t}[-\log p_{T}(x_t |z_t,r_t)]
  \label{eq5}
\end{split}
\end{equation}
where $\theta$ denotes the parameter of the corresponding module. In this way, we can maintain reconstruction fidelity of an utterance with the synthesizer $T$.

Besides, for a sample $x_s$, we enforce the decoded sequence, given its speaker representation $r_s$ and target style representation $z_t$, should be in the target domain $X_t$. Following GAN~\cite{Ian2014generative, mathieu2016disentangling}, we introduce an adversarial loss $\mathcal{L}_{adv}$ to be minimized in decoding and adopt a discriminator $D$, as shown in Fig.~\ref{fig1}, to distinguish from $T(r_s, z_t)$ and $T(r_t, z_t)$. The task of the synthesizer is to fool the discriminator. Specifically, the adversarial loss $\mathcal{L}_{adv}$ is defined as
\begin{equation}
\begin{split}
   &\mathcal{L}_{adv}(\theta_{E_r},\theta_{E_z},\theta_{T},\theta_{D})\\
   &=\mathbb{E}_{x_s \sim X_s}[-\log (1-D(T(r_s,z_t)))]\\
   &+\mathbb{E}_{x_t \sim X_t}[-\log D(T(r_t,z_t))]
  \label{eq6}
\end{split}
\end{equation}

But, for a sample $x_s \in X_s$, its $z_s$ can be an arbitrary value that minimizes the above reconstruction loss and adversarial loss, which may not necessarily capture the utterance style. This will affect the speaker representation, which is critical for representing the speaker, and it should be invariant against the transferred style. To address the issue, we introduce a style distortion loss $\mathcal{L}_{dis}$ to constrain style representation of a source utterance to be closer to the target style representation. As shown in Fig.~\ref{fig1}, a discriminator, $D_s$, is first trained to predict whether a given utterance $x$ has the target style with an output probability, $p_{Ds}(x \in X_t)$. When learning the style representation $z_s$, we then force the distortion between this style representation $z_s$ and target style representation $z_t$ to be consistent with the output probability of $D_s$. Here, we use the $L_2$ norm to measure the style distortion, $d(z_s, z_t)=\|z_s -z_t\|_2$, and make the style distortion positively correlated with $1-p_{Ds}(x_s \in X_t)$. To incorporate this idea into our model, we model it with a standard normal distribution to evaluate the style distortion loss. Intuitively, when an utterance $x_s$ have a large output probability $p_{Ds}(x_s \in X_t)$, our model can result in a small style distortion. That is, $z_s$ will be closer to $z_t$, and the style distortion loss $\mathcal{L}_{dis}$ is:
\begin{equation}
\begin{split}
   &\mathcal{L}_{dis}(\theta_{E_z})=\mathbb{E}_{x_s \sim X_s}[p_{Ds}(x_s \in X_t)d(z_s, z_t)^2]
  \label{eq7}
\end{split}
\end{equation}
where $D_s$ is a pre-trained model trained with a portion of the training data. Here, if we integrate $D_s$ into our training, we may start with a $D_s$ with a low accuracy, and then our model is inclined to optimize a wrong style distortion loss for many epochs and gets stuck into a poor local optimum.

Unfortunately, the style distortion loss $\mathcal{L}_{dis}$ can only constrain the generated utterance to be aligned with the target style, but cannot guarantee to keep the speaker identity of a source utterance intact. To address this issue, we introduce a cycle consistency loss~\cite{zhu2017unpaired, Liumeng2021on}, $\mathcal{L}_{cyc}$, to our model shown in Fig.~\ref{fig1}, which requires a transferred utterance to preserve the speaker identity of its source utterance, and enables the recovery of the source utterance in a cyclic manner. The cycle consistency loss $\mathcal{L}_{cyc}$ is defined as
\begin{equation}
\begin{split}
   &\mathcal{L}_{cyc}(\theta_{E_r},\theta_{E_z},\theta_{T})\\
   &=\mathbb{E}_{x_s \sim X_s}[-\log p_{T}(x_s| E_r (\widetilde{x}_s), z_s)]\\
   &+\mathbb{E}_{x_t \sim X_t}[-\log p_{T}(x_t| E_r (\widetilde{x}_t), z_t)]
  \label{eq8}
\end{split}
\end{equation}
where $\widetilde{x}_s$ is the transferred utterance from a source sample $x_s$ and has the target style $z_t$. We encode the $\widetilde{x}_s$ with the speaker encoder $E_r(\widetilde{x}_s)$ to obtain its speaker representation $\widetilde{r}_s$, which is then combined with its source style $z_s$ for decoding. Here, we expect that the source utterance can be generated with a high probability. For a target sample $x_t$, although we do not aim at changing its style in our model, similar to $x_s$, we still calculate its cycle consistency loss for additional regularization.

In addition, to learn more discriminative style and speaker representations that can distinguish different styles and capture the characteristics of different speakers, we introduce a style classification loss $\mathcal{L}_{stycls}$ and a speaker classification loss $\mathcal{L}_{spkcls}$, as shown in Fig.~\ref{fig1}. In this paper, we let the style classifier and the speaker classifier have the same architecture, and train the two classifiers with the cross-entropy loss, or the softmax loss. The details of the two classifiers will be described in Section~\ref{Network}. Specially, the softmax loss $\mathcal{L}_{softmax}$ (i.\,e., the style classification loss $\mathcal{L}_{stycls}$ and the speaker classification loss $\mathcal{L}_{spkcls}$) is formulated as
\begin{equation}
	\mathcal{L}_{softmax}=-\sum_{i,j}y_{i,j}\log(\hat{y}_{i,j})
	\label{eq9}
	\end{equation}
where $i$ refers to the style/speaker index of the input style/speaker embedding and $j$ refers to the style/speaker index upon which the classification occurs. The $y_{i,j}$ is the ground-truth style/speaker class for the $i$-th embedding in the $j$-th style/speaker dimension, and $\hat{y}_{i,j}$ is the predicted style/speaker class. For $i=j$, the two classifiers encourage a style/speaker embedding to contain the correct information of the $i$-th style/speaker. For $i\neq j$, the two classifiers discourage the use of information about the other style/speaker.
	
	By joint-training to minimize the weighted six specific objectives, or the overall objective function $\mathcal{L}$
	\begin{equation}
    \begin{split}
	&\mathcal{L}=\alpha\mathcal{L}_{rec}+\beta\mathcal{L}_{adv}+\gamma\mathcal{L}_{dis}\\
    &+\lambda\mathcal{L}_{cyc}+\kappa\mathcal{L}_{stycls}+\omega\mathcal{L}_{spkcls}
	\label{eq10}
    \end{split}
	\end{equation}
	where $\alpha/\beta/\gamma/\lambda/\kappa/\omega$ are coefficients of the reconstruction term $\mathcal{L}_{rec}$ in Eq.(\ref{eq5}), the adversarial loss $\mathcal{L}_{adv}$ in Eq.(\ref{eq6}), the style distortion loss $\mathcal{L}_{dis}$ in Eq.(\ref{eq7}), the cycle consistency loss $\mathcal{L}_{cyc}$ in Eq.(\ref{eq8}), the style classification loss $\mathcal{L}_{stycls}$ and the speaker classification loss $\mathcal{L}_{spkcls}$ in Eq.(\ref{eq9}), respectively.
		
	\section{Experimental Setup}
	\label{ExperimentalSetup}
	\subsection{Datasets and Comparison Models}
	\label{Datasets}	
	We have carried out experiments to evaluate the performance of the proposed approach. We focus on disjoint, multi-style datasets, where datasets of different styles are recorded, and each style is recorded by one speaker with multiple utterances. Here, an internal Chinese corpus is used in experiments, which is divided into source data and target data. Source data contains examples of four styles: \textit{reading} (standard reading speech), \textit{broadcasting} (news broadcasting speech), \textit{talking} (conversational speech) and \textit{story-telling} (audiobook speech) styles, from four different speakers whereas target data contains samples of three styles: \textit{customer-service} (spontaneous speech with fast speed), \textit{poetry} (classical Chinese poetry speech with rich prosody variations), and \textit{game} (exaggerated speech for role dubbing in the game) styles, from three other different speakers. This represents a minimalistic scenario of the disjoint, multi-style datasets: a single model must be able to properly transfer an arbitrary or unknown style to target style while keeping a minimal perceived change in the speaker's timbre. The corpus contains 22,212 samples ($\sim$ 25 hours) and each style contains 4,000 samples except for poetry and game styles. There are 698 samples in poetry style while 1514 samples in game style. For samples of each style, we respectively use 90\% as the training set, 5\% as the validation set and the rest 5\% for the test. We remove long silence ($\textgreater$ 0.1 sec) at the beginning and ending of each utterance. 80-dim Mel spectrogram is extracted as target speech representations with a Hanning window of 50 ms and 12.5 ms frame shift. Phoneme sequences are used as the input, which are obtained by text normalization and G2P pre-processing modules.

For all different systems in our experiments, we train $\sim$ 220k steps with a single Nvidia Tesla P40 GPU, and a batch size is 32. The models trained and tested in our experiments include:
\begin{itemize}
  \item GST: we introduce ``global style tokens" (GST)~\cite{wang2018style} into Tacotron 2~\cite{shen2018natural} to uncover expressive factors of variation in speaking style to perform the style transfer task, and make a fair comparison;
  \item VAE: we incorporate VAE~\cite{zhang2019learning} into Tacotron 2~\cite{shen2018natural} to learn the latent representation of speaking style to guide the style in synthesizing speech;
  \item MRF-IT: we augment a multi-reference encoder into GST-Tacotron 2~\cite{bian2019multi} to model multiple styles simultaneously, and adopt intercross training to extract and separate different classes of speech styles, thus achieving the transfer for desired speech styles;
  \item MRF-ACC: we adopt an adversarial cycle consistency training scheme for multi-reference neural TTS stylization~\cite{whitehill2020multi} to ensure the use of information from all style classes, thus performing multi-reference style transfer on disjoint datasets;
  \item Proposed model: we introduce an IAF technique~\cite{kingma2016improving} to improve variational inference and learn expressive style representations, and develop a joint-training speaker encoder network to obtain discriminative speaker representations. Six loss functions with different purposes in network training are used together for enhancing the performance of seen and unseen style transfer on disjoint, multi-style datasets.
\end{itemize}
			
		\begin{table*}[!t]
			\caption{\label{tab:naturalness} {MOS results with 95 \% confidence interval of seen and unseen style transfer from different models.}}
			\centerline{
            %\scalebox{0.75}{
				\begin{tabular}{|c|ccc|ccc|}
					\hline
					\multirow{2}*{System} & \multicolumn{3}{c|}{\underline{Seen style transfer}} & \multicolumn{3}{c|}{\underline{Unseen style transfer}} \\
					   & {R2C} & {R2P}  & {R2G}   & {TR2C} & {TR2P}  & {TR2G}  \\
					\hline
					 GST     &3.35$\pm$0.04    &3.27$\pm$0.02  & 3.24$\pm$0.01   &2.92$\pm$0.02  &2.83$\pm$0.03  & 2.81$\pm$0.05    \\
					 VAE     &3.43$\pm$0.02    &3.39$\pm$0.04  & 3.35$\pm$0.02   &2.97$\pm$0.03  &2.92$\pm$0.07  & 2.89$\pm$0.06    \\			
					MRF-IT   &3.56$\pm$0.05    &3.49$\pm$0.06  & 3.45$\pm$0.07   &3.09$\pm$0.05  &3.01$\pm$0.12  & 2.97$\pm$0.08    \\
					MRF-ACC  &3.78$\pm$0.03    &3.70$\pm$0.04  & 3.63$\pm$0.06   &3.58$\pm$0.04  &3.49$\pm$0.04  & 3.47$\pm$0.05    \\
					Proposed &\textbf{3.99$\pm$0.01} &\textbf{3.95$\pm$0.02} &\textbf{3.91$\pm$0.01}  &\textbf{3.86$\pm$0.02} &\textbf{3.83$\pm$0.03} &\textbf{3.81$\pm$0.02} \\
					\hline
				\end{tabular}
			}
           % }
		\end{table*}

		\begin{table*}[!t]
			\caption{\label{tab:ABXnaturalness} {ABX preference results between the proposed model and each reference system for speech naturalness on seen and unseen style transfer.}}
			\centerline{
				\begin{tabular}{|c|ccc|ccc|ccc|}
					\hline
                      \multicolumn{10}{|c|}{Preference (\%) for seen style transfer}  \\
                    \hline
					\multirow{2}*{System A vs System B} &\multicolumn{3}{c|}{\underline{R2C}} & \multicolumn{3}{c|}{\underline{R2P}} &\multicolumn{3}{c|}{\underline{R2G}} \\
                      &{System A} & {Neutral}  & {System B}  & {System A} & {Neutral} &{System B} & {System A} & {Neutral}  & {System B} \\
					\hline
					  GST vs Proposed    &28.5 &31.2 &\textbf{40.3} &28.9 &31.0 &\textbf{40.1} &29.0 &31.2 &\textbf{39.8} \\
					  VAE vs Proposed    &29.4 &31.0 &\textbf{39.6} &30.1 &30.1 &\textbf{39.8} &30.7 &29.9 &\textbf{39.4} \\			
					  MRF-IT vs Proposed &31.2 &29.1 &\textbf{39.7} &31.5 &29.3 &\textbf{39.2} &31.9 &29.1 &\textbf{39.0} \\
				      MRF-ACC vs Proposed&33.5 &27.5 &\textbf{39.0} &33.3 &28.1 &\textbf{38.6} &33.5 &28.1 &\textbf{38.4} \\
					\hline
                      \multicolumn{10}{|c|}{Preference (\%) for unseen style transfer}  \\
                    \hline
					\multirow{2}*{System A vs System B} &\multicolumn{3}{c|}{\underline{TR2C}} & \multicolumn{3}{c|}{\underline{TR2P}} &\multicolumn{3}{c|}{\underline{TR2G}} \\
                      &{System A} & {Neutral}  & {System B}  & {System A} & {Neutral} &{System B} & {System A} & {Neutral}  & {System B} \\
					\hline
					  GST vs Proposed    &24.9 &31.9 &\textbf{43.2} &25.2 &32.5 &\textbf{42.3} &25.8 &32.3 &\textbf{41.9} \\
					  VAE vs Proposed    &25.6 &33.1 &\textbf{41.3} &26.1 &33.0 &\textbf{40.9} &27.0 &31.9 &\textbf{41.1} \\			
					  MRF-IT vs Proposed &29.6 &29.7 &\textbf{40.7} &29.4 &30.2 &\textbf{40.4} &29.2 &30.8 &\textbf{40.0} \\
				      MRF-ACC vs Proposed&31.8 &29.0 &\textbf{39.2} &31.5 &29.0 &\textbf{39.5} &31.7 &28.7 &\textbf{39.6} \\
					\hline
				\end{tabular}
			}
		\end{table*}

	\subsection{Model Details}
	\label{Network}
	The style encoder contains a reference encoder, an IAF flow, and a style classifier, which forms a more discriminative and expressive latent style representation. Similar to ~\cite{skerry2018towards, zhang2019learning}, the reference encoder consists of a stack of six 2-D convolutional layers cascaded with one unidirectional 128-unit GRU layer. In each IAF step, we use the structure proposed in ~\cite{germain2015made} as the autoregressive neural network. As shown in Fig.~\ref{fig2}, the style classifier contains three FC layers, all with ReLU activation, and a softmax output layer. We use the output of the third FC layer in the style classifier as the style embeddings. For the speaker encoder, we use a 3-layer LSTM with the projection operations followed by a speaker classifier shown in Fig.~\ref{fig4}. In this paper, we let the style classifier and the speaker classifiers have the same architecture, to learn discriminative style representations and speaker embeddings that can make both style and speaker to be more distinguishable.

In our model, we adopt Tacotron 2~\cite{shen2018natural} as the synthesizer, which takes the concatenation of the speaker and style representations as the initial hidden state. As for the discriminator $D$, we follow the architecture of the discriminator in ~\cite{guo2019new}. The pre-trained discriminator $D_s$ used in the style distortion loss has the same structure as the style encoder followed by a sigmoid output layer. As for the objective function in Eq.(\ref{eq10}), we empirically set the coefficients $\alpha$, $\beta$, $\lambda$, $\kappa$ and $\omega$ to 1.0, and $\gamma$ to 5.0, respectively. The six coefficients ($\alpha/\beta/\gamma/\lambda/\kappa/\omega$) are preset weights for balancing the different loss terms ($\mathcal{L}_{rec}$/$\mathcal{L}_{adv}$/$\mathcal{L}_{dis}$/$\mathcal{L}_{cyc}$/$\mathcal{L}_{stycls}$/$\mathcal{L}_{spkcls}$).

In this paper, we use LPCNet, a variant of WaveRNN~\cite{kalchbrenner2018efficient}, as a neural vocoder to convert the rendered Mel spectrograms by the synthesizer network into time speech waveforms. The architecture is the same as that described in ~\cite{Valin2019LPCNET}, composed of a sample rate network and a frame rate network. Different from ~\cite{Valin2019LPCNET}, we adopt the Mel spectrogram as the input, rather than the Bark-scale cepstral coefficients~\cite{moore2012introduction} and pitch parameters, thus performing a direct connection with the synthesizer network. The training pipeline of our LPCNet is the same as that of ~\cite{Valin2019LPCNET}. Here, we use the same LPCNet for fair comparisons across all different systems.

	\subsection{Evaluation}
	The performance of seen and unseen speech style transfer is evaluated in speech naturalness, style similarity, and speaker similarity. We conduct Mean Opinion Score (MOS) and preference listening tests (ABX) of speech naturalness to evaluate the reconstruction performance of different experimental systems. An ABX test of style similarity is also conducted to assess the style conversion performance, where subjects are asked to choose which speech sample (A or B) sounds closer to the target style (X) in terms of style. We further conduct a Comparative Mean Opinion Score (CMOS) test of speaker similarity to evaluate how well the transferred speech matches that of the source speaker. For each system, we randomly select the reading style as the seen style and make three experiments: from Reading style to Customer-service style (R2C), Reading style to Poetry style (R2P), and Reading style to Game style (R2G). We randomly choose an unique Taiwanese-reading style from a new female speaker as the unseen style, and conduct three tests from Taiwanese-reading style to Customer-service style (TR2C), Taiwanese-reading style to Poetry style (TR2P), and Taiwanese-reading style to Game style (TR2G) to assess the performance of unseen style transfer.

In addition, style classification accuracy and visualization of style embedding space are adopted to further evaluate the style similarity. Speaker classification accuracy and cosine similarity are calculated to measure the speaker similarity objectively.

	\section{Experimental Results}
	\label{ExperimentalResults}

	\subsection{Speech Naturalness}
	\label{Speech naturalness}		
	We use the seen and unseen style evaluation sets, which respectively contain 20 sentences in each style, randomly selected from the test set, to compare the performance of all models in speech naturalness with the MOS and ABX listening tests\footnote{Samples can be found at \url{https://xiaochunan.github.io/disentangling/index.html}}. Table~\ref{tab:naturalness} summarizes the results of MOS from different models, where 15 subjects are requested to carefully listen and evaluate with rating scores from 1 to 5 in 0.5 point increments. It can be seen that our proposed approach outperforms the reference models on both seen and unseen style transfer tasks. The performance of the proposed model on unseen style transfer is much better than other models. Specially, most of subjects find that there are more unintelligible parts in an utterance synthesized by using the GST, VAE, MRF-IT and MRF-ACC models. This phenomenon is more obvious on unseen style transfer task. The results show a better generalization of the proposed model on the unseen style transfer. Seen style transfer performs with better speech naturalness than the unseen style transfer. Partially due to the small size of the speech data set in poetry and game styles, hence their MOS scores are slightly lower than that of the customer-service style.

Table~\ref{tab:ABXnaturalness} shows the results of ABX tests between the proposed model and all reference systems for speech naturalness on seen and unseen style transfer, where the same 15 subjects are asked to choose the speech samples with higher speech naturalness. We can observe that our proposed framework consistently outperforms four other systems of the prior art, which is also consistent with the MOS results. These observations verify the effectiveness of the proposed approach in terms of speech naturalness.

We also calculate the character error rate (CER) of all models on the same evaluation sets via a Conformer based ASR system~\cite{guo2021recent} to evaluate speech intelligibility objectively. As shown in Table~\ref{tab:cer}, we show that our proposed model is superior to the four reference models, i.e., GST, VAE, MRF-IT and MRF-ACC, on both seen and unseen style transfer. This verifies objectively the effectiveness of our proposed approach in terms of speech intelligibility.

\begin{table}[!t]
			\caption{\label{tab:cer} {CER results (\%) of different models on seen and unseen style transfer.}}
			\centerline{
            %\scalebox{0.75}{
				\begin{tabular}{|c|ccc|ccc|}
					\hline
					\multirow{2}*{System} & \multicolumn{3}{c|}{\underline{Seen style transfer}} & \multicolumn{
3}{c|}{\underline{Unseen style transfer}} \\
					   & {R2C} & {R2P}  & {R2G}   & {TR2C} & {TR2P}  & {TR2G}  \\
					\hline
					 GST     &20.5    &20.7  & 20.8   &22.8  &23.4  & 23.7    \\
					 VAE     &19.9    &20.3  & 20.5   &22.4  &22.9  & 23.1    \\			
					MRF-IT   &19.2    &19.6  & 19.7   &21.9  &22.1  & 22.3    \\
					MRF-ACC  &17.8    &18.2  & 18.6   &19.1  &19.4  & 19.5    \\
					Proposed &\textbf{15.4} &\textbf{15.6} &\textbf{15.8}  &\textbf{16.2} &\textbf{16.4} &\textbf{16.7} \\
					\hline
				\end{tabular}
			}
           % }
		\end{table}

\begin{table}[!t]
			\caption{\label{tab:styleclassification} {Results of style classification (\%) for seen and unseen style transfer using different models.}}
			\centerline{
            %\scalebox{0.75}{
				\begin{tabular}{|c|ccc|ccc|}
					\hline
					\multirow{2}*{System} & \multicolumn{3}{c|}{\underline{Seen style transfer}} & \multicolumn{
3}{c|}{\underline{Unseen style transfer}} \\
					   & {R2C} & {R2P}  & {R2G}   & {TR2C} & {TR2P}  & {TR2G}  \\
					\hline
					 GST     &62.8    &62.2  & 61.5   &58.7  &57.6  & 56.2    \\
					 VAE     &64.2    &63.8  & 63.0   &59.9  &59.2  & 58.4    \\			
					MRF-IT   &71.3    &70.4  & 69.2   &67.7  &67.1  & 66.5    \\
					MRF-ACC  &76.8    &76.3  & 75.4   &73.2  &72.5  & 71.6    \\
					Proposed &\textbf{90.5} &\textbf{90.2} &\textbf{89.9}  &\textbf{87.6} &\textbf{86.9} &\textbf{85.8} \\
					\hline
				\end{tabular}
			}
           % }
		\end{table}

\begin{figure*}[!t]
  \centering
  \includegraphics[width=0.82\linewidth]{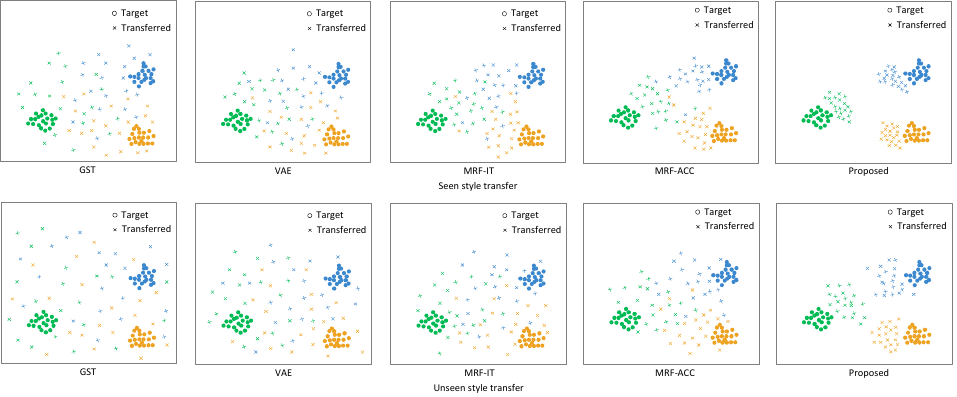}
  \caption{t-SNE plots of style embeddings for seen and unseen style transfer using different models. Each color corresponds to a different style. Real and transferred utterances appear nearby when they are from the same style, however real and transferred utterances consistently form distinct clusters.}
  \label{fig:tsnestyleembedding}

\end{figure*}
		\begin{table*}[!t]
			\caption{\label{tab:stylesimilarityabx} {ABX preference results between the proposed model and each reference system for style similarity on seen and unseen style transfer.}}
			\centerline{
				\begin{tabular}{|c|ccc|ccc|ccc|}
					\hline
                      \multicolumn{10}{|c|}{Preference (\%) for seen style transfer}  \\
                    \hline
					\multirow{2}*{System A vs System B} &\multicolumn{3}{c|}{\underline{R2C}} & \multicolumn{3}{c|}{\underline{R2P}} &\multicolumn{3}{c|}{\underline{R2G}} \\
                      &{System A} & {Neutral}  & {System B}  & {System A} & {Neutral} &{System B} & {System A} & {Neutral}  & {System B} \\
					\hline
					  GST vs Proposed    &26.8 &32.9 &\textbf{40.3} &27.2 &32.0 &\textbf{40.8} &27.6 &32.1 &\textbf{40.3} \\
					  VAE vs Proposed    &27.7 &31.7 &\textbf{40.6} &28.1 &31.7 &\textbf{40.2} &28.5 &31.5 &\textbf{40.0} \\			
					  MRF-IT vs Proposed &28.9 &31.2 &\textbf{39.9} &29.3 &30.6 &\textbf{40.1} &29.4 &30.7 &\textbf{39.9} \\
				      MRF-ACC vs Proposed&29.6 &31.2 &\textbf{39.2} &30.2 &30.0 &\textbf{39.8} &31.4 &29.1 &\textbf{39.5} \\
					\hline
                      \multicolumn{10}{|c|}{Preference (\%) for unseen style transfer}  \\
                    \hline
					\multirow{2}*{System A vs System B} &\multicolumn{3}{c|}{\underline{TR2C}} & \multicolumn{3}{c|}{\underline{TR2P}} &\multicolumn{3}{c|}{\underline{TR2G}} \\
                      &{System A} & {Neutral}  & {System B}  & {System A} & {Neutral} &{System B} & {System A} & {Neutral}  & {System B} \\
					\hline
					  GST vs Proposed    &24.8 &33.2 &\textbf{42.0} &25.3 &32.9 &\textbf{41.8} &25.7 &32.4 &\textbf{41.9} \\
					  VAE vs Proposed    &25.7 &32.0 &\textbf{42.3} &26.1 &31.9 &\textbf{42.0} &26.3 &32.1 &\textbf{41.6} \\			
					  MRF-IT vs Proposed &26.4 &32.3 &\textbf{41.3} &27.0 &31.6 &\textbf{41.4} &27.6 &31.7 &\textbf{40.7} \\
				      MRF-ACC vs Proposed&27.5 &31.7 &\textbf{40.8} &28.2 &29.9 &\textbf{41.9} &28.7 &30.8 &\textbf{40.5} \\
					\hline
				\end{tabular}
			}
		\end{table*}
	\subsection{Style Similarity}
	\label{Style similarity}
	 We first investigate the style classification accuracy via a speech style classifier, which is independently trained using the seven style samples (i.\,e., reading, broadcasting, talking, story-telling, customer-service, poetry and game styles) from the training data, to objectively evaluate the style conversion performance. The classifier has the same architecture as the style encoder in our model. Its final validation accuracy is 95.1\%. We then synthesize the transferred speech adopting the same evaluation sets, and predict their style labels using the trained classifier. Table~\ref{tab:styleclassification} shows the results of style classification for seen and unseen style transfer using different models. Our proposed model achieves greater accuracy on both seen and unseen style transfer tasks, showing its ability to transfer style in synthesized samples. However, the reference models perform poorly on style classification. Specially, for the unseen style transfer (i.\,e., TR2C, TR2P, and TR2G), many transferred samples from the GST, VAE, and MRF-IT models are difficult to distinguish as their samples after the transfer always have poor style expression, demonstrating a poor unseen style transfer performance. The best reference system, MRF-ACC, still obtains a much lower rate of seen and unseen style transfer than the proposed model. These results validate the effectiveness of the proposed approach for seen and unseen style transfer.

We further visualize the style embedding space using t-SNE~\cite{van2008visualizing} to evaluate the style similarity, where different colors correspond to different styles. The style embeddings are respectively extracted from 20 real target speech and 20 transferred utterances for each style. As shown in Fig.~\ref{fig:tsnestyleembedding}, our proposed model produces much closer and more separable clusters than the other four reference systems. Particularly, in our model, different styles are well separated from each other in the style embedding space, and transferred utterances tend to lie close to real target speech from the same style in the embedding space. However, the transferred utterances are still easily distinguishable from the real human speech as demonstrated by the t-SNE visualization, where utterances from each transferred style form a nice cluster which is adjacent to a cluster of real target utterances of the corresponding style. For the GST, VAE, MRF-IT, and MRF-ACC models, real sentences form distinct clusters, but transferred utterances cannot be well separated by style, and appear distant from the corresponding real target style utterances. These observations suggest that our approach can learn effective and discriminative representations in the style space, thus enhancing the performance of both seen and unseen style transfer.

We also conduct an ABX test of style similarity between the proposed model and each reference system to subjectively assess the style conversion performance, where the same 15 listeners are asked to choose the speech samples which sound closer to the target style in terms of style expression. Here, the listeners follow the instructions: ``You should not judge the content, grammar, audio quality, or speaker identity of the sentences; instead, just focus on the similarity of the style to one another.". Higher preference means more style similarity is perceived. The results of style similarity on seen and unseen style transfer are shown in Table~\ref{tab:stylesimilarityabx}, where the listeners give higher preference to the proposed system, showing the proposed approach improves the performance of seen and unseen style transfer. For the unseen style transfer, we find that the GST, VAE and MRF-IT models, in most cases, fail to transfer unseen style of the Taiwanese-reading style to the target style of customer-service style or poetry style or game style. The MRF-ACC system, the best of all references, is still significantly inferior to the proposed model in the style similarity test. These results demonstrate the effectiveness of our proposed approach for both seen and unseen style transfer.

	\subsection{Speaker Similarity}	
	\label{Speaker similarity}
To evaluate how well the transferred speech matches that of the source speaker's timbre, we respectively conduct CMOS tests between the proposed model and each reference system (i.\,e., proposed vs GST, proposed vs VAE, proposed vs MRF-IT, and proposed vs MRF-ACC) by using the same evaluation sets. Here, CMOS is used to make a comparison in speaker similarity between two voices, ranging from -3 to 3. Generally, a positive score indicates the reference voice is better than the proposed voice in speaker similarity, worse for a negative score. The same 15 listeners are asked to just focus on the speaker identity of utterances while not judging the content, grammar, audio quality, or style information of the utterances. Table~\ref{tab:cmosspeakersimilarity} reports CMOS results for speaker similarity, where score of the proposed model is fixed to 0 on both seen and unseen style transfer tasks. We can observe that each reference system obtains negative CMOS scores, demonstrating that they perform worse in speaker similarity than the proposed model. This is also evident that our proposed approach delivers better speaker similarity performance than all reference systems, on both seen and unseen style transfer tasks.

We also adopt the cosine similarity to calculate the similarity between the speaker embedding of a transferred sample and the speaker embedding of a randomly selected ground truth utterance from the same speaker to objectively measure the speaker similarity. The results are shown in Table~\ref{tab:cosinedistance}, where we can find that the proposed model delivers a higher speaker similarity than the other reference systems, reflecting that our approach has learned a more discriminative speaker representation. Specially, our proposed model obtains a greater cosine similarity, which demonstrates that the transferred utterances via our method are nearly always even closer to the speaker identity of source speaker. On the unseen style transfer of TR2C, TR2P and TR2G, the GST, VAE and MRF-IT models are not capable of keeping the speaker's timbre, resulting in lower speaker similarity. The best reference system, MRF-ACC, still performs significantly worse than the proposed approach.

We further train a speaker verification model to calculate speaker classification accuracy for the same evaluation sets to objectively assess the speaker similarity. In this experiments, we follow the architecture of speaker verification model in ~\cite{zhang2021multi}, which consists of a small size Thin ResNet-34~\cite{heo2020clova} with SE-block~\cite{zhou2019deep}. Table~\ref{tab:speakerclassification} summarizes the results of speaker classification accuracy on both seen and unseen style transfer tasks. Our proposed approach obtains a higher speaker classification accuracy than the other reference systems. Specially, on the unseen style transfer task (i.\,e., TR2C, TR2P, and TR2G), the proposed model achieves a much higher accuracy than the GST, VAE, and MRF-IT models, showing its ability to retain speaker identity in the transferred samples. The MRF-ACC system, the best reference model, is still significantly inferior to the proposed method in the speaker classification test. From these results we can conclude that our proposed model can preserve speaker identity and generate speech that resembles the source speaker, but still not as good as the real source speaker.

\begin{table}[!t]
			\caption{\label{tab:cmosspeakersimilarity} {CMOS results for speaker similarity between the proposed model and each reference system.}}
			\centerline{
            %\scalebox{0.75}{
				\begin{tabular}{|c|ccc|ccc|}
					\hline
					 \multirow{2}*{System}& \multicolumn{3}{c|}{\underline{Seen style transfer}} & \multicolumn{
3}{c|}{\underline{Unseen style transfer}} \\
					   & {R2C} & {R2P}  & {R2G}   & {TR2C} & {TR2P}  & {TR2G}  \\
					\hline
					Proposed     &0        &0      & 0       &0      &0      &0    \\
					 GST         &-0.83    &-0.85  & -0.86   &-0.94  &-0.95  & -0.97    \\			
					 VAE         &-0.70    &-0.68  & -0.67   &-0.76  &-0.78  & -0.80    \\
					MRF-IT       &-0.45    &-0.48  & -0.49   &-0.52  &-0.53  & -0.56    \\
					MRF-ACC      &-0.22    &-0.24  & -0.25   &-0.30  &-0.33  & -0.34 \\
					\hline
				\end{tabular}
			}
           % }
		\end{table}

\begin{table}[!t]
			\caption{\label{tab:cosinedistance} {Cosine similarity results for speaker similarity using different models.}}
			\centerline{
            %\scalebox{0.75}{
				\begin{tabular}{|c|ccc|ccc|}
					\hline
					\multirow{2}*{System} & \multicolumn{3}{c|}{\underline{Seen style transfer}} & \multicolumn{
3}{c|}{\underline{Unseen style transfer}} \\
					   & {R2C} & {R2P}  & {R2G}   & {TR2C} & {TR2P}  & {TR2G}  \\
					\hline
					 GST     &0.30    &0.27  & 0.26   &0.20  &0.18  & 0.17    \\
					 VAE     &0.36    &0.34  & 0.33   &0.24  &0.23  & 0.21    \\			
					MRF-IT   &0.44    &0.43  & 0.41   &0.35  &0.34  & 0.33    \\
					MRF-ACC  &0.58    &0.56  & 0.53   &0.44  &0.42  & 0.41    \\
					Proposed &\textbf{0.78} &\textbf{0.76} &\textbf{0.75}  &\textbf{0.71} &\textbf{0.70} &\textbf{0.68} \\
					\hline
				\end{tabular}
			}
           % }
		\end{table}

\begin{table}[!t]
			\caption{\label{tab:speakerclassification} {Results of speaker classification (\%) for seen and unseen style transfer using different models.}}
			\centerline{
            %\scalebox{0.75}{
				\begin{tabular}{|c|ccc|ccc|}
                    \hline
					\multirow{2}*{System} & \multicolumn{3}{c|}{\underline{Seen style transfer}} & \multicolumn{
3}{c|}{\underline{Unseen style transfer}} \\
					   & {R2C} & {R2P}  & {R2G}   & {TR2C} & {TR2P}  & {TR2G}  \\
					\hline
					 GST     &62.9    &62.4  & 61.5   &59.8  &59.1  & 58.0    \\
					 VAE     &64.7    &63.6  & 63.0   &61.9  &60.5  & 60.3    \\			
					MRF-IT   &71.4    &70.8  & 70.1   &68.2  &67.4  & 67.1    \\
					MRF-ACC  &78.2    &78.0  & 77.3   &75.6  &74.8  & 73.5    \\
					Proposed &\textbf{92.1} &\textbf{91.6} &\textbf{91.2}  &\textbf{90.3} &\textbf{90.0} &\textbf{89.4} \\
					\hline
				\end{tabular}
			}
           % }
		\end{table}

\begin{figure}[!t]
  \centering
  \includegraphics[width=0.85\linewidth]{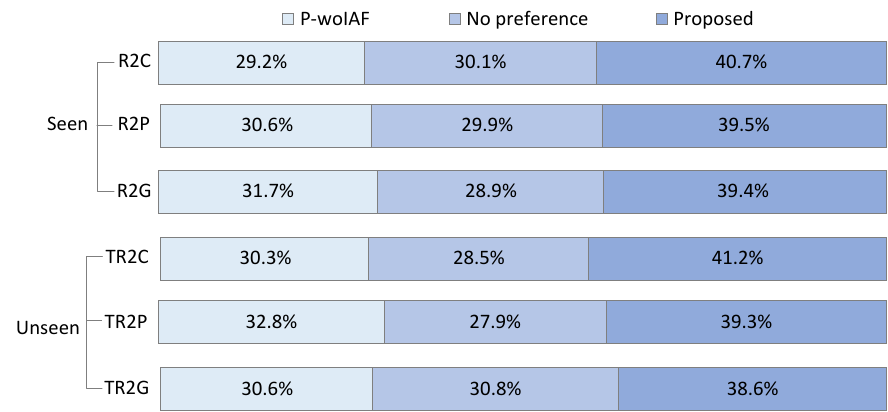}
  \caption{ABX preference results between P-woIAF and proposed model for style similarity on seen and unseen style transfer.}
  \label{fig:IAFstylesimilarity}
\end{figure}

	\subsection{Investigation on Style Transfer without IAF}	
	\label{Investigation on style transfer without IAF}
	It is meaningful to investigate the effect of not using IAF in learning style representations for speech style transfer. In this experiments, we remove the IAF structure in the style encoder and test the performance of its seen and unseen style transfer, which is denoted as P-woIAF. Adopting the same evaluation sets, we conduct an ABX test of style similarity between the P-woIAF and the proposed approach to subjectively compare their style conversion performance. Here the same 15 listeners are also asked to choose the speech samples that sound closer to the target style in terms of style expression. Fig.~\ref{fig:IAFstylesimilarity} presents the ABX results for style similarity on both seen and unseen style transfer tasks. We can find that the P-woIAF model obtains lower preference, showing lower style similarity is perceived. However, when we plug the IAF to the style encoder, i.\,e., our proposed model, the listeners give higher preference to the proposed approach. The performance gain is essentially contributed by the IAF scheme in learning expressive style representations. In summary, we should construct sophisticated distributions with IAF to learn discriminative style representations, for possible improved performance in seen and unseen style transfer.  	
\begin{figure}[!t]
  \centering
  \includegraphics[width=0.9\linewidth]{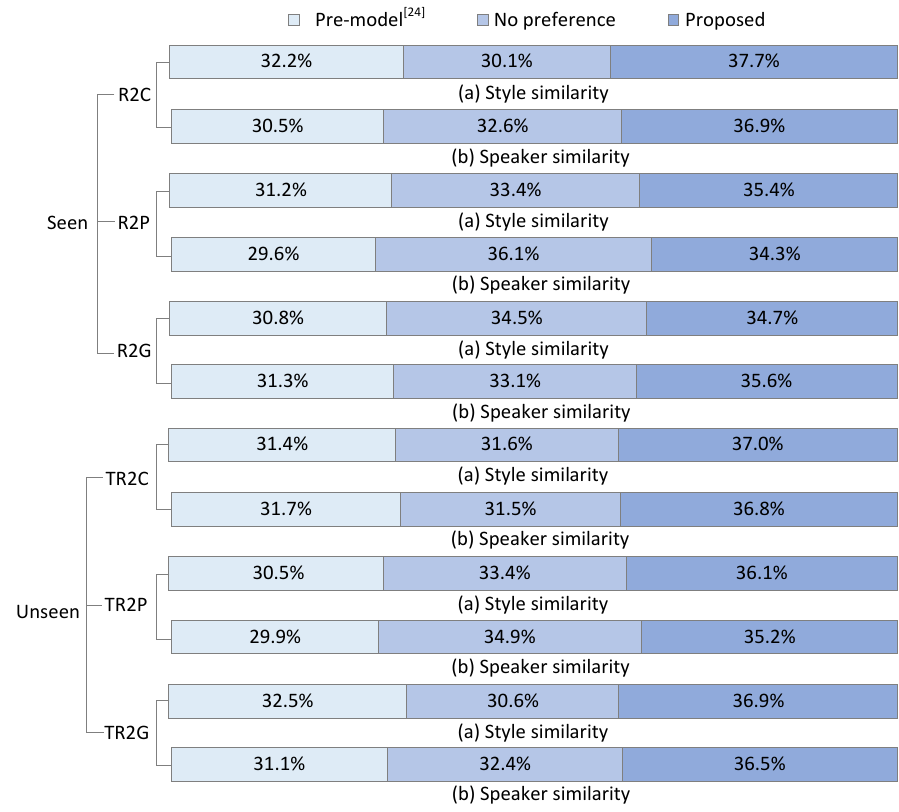}
  \caption{ABX preference results between Pre-model\textsuperscript{\cite{An2021improving}} and proposed model for style similarity and speaker similarity on seen and unseen style transfer.}
  \label{fig:withoutclassifierabx}
\end{figure}

\begin{figure}[!t]
  \centering
  \includegraphics[width=1.0\linewidth]{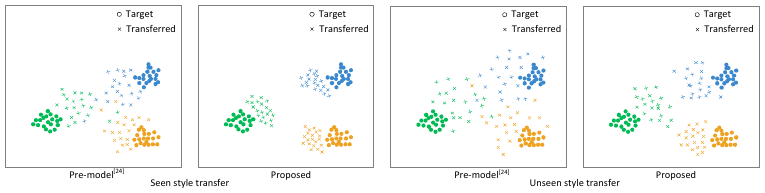}
  \caption{t-SNE plots of style embeddings for Pre-model\textsuperscript{[24]} and proposed model on seen and unseen style transfer. Each color corresponds to a different style. Real and transferred utterances appear nearby when they are from the same style, however real and transferred utterances consistently form distinct clusters.}
  \label{fig:tsneclassifers}
\end{figure}

	\subsection{Comparison of Style Transfer without Style and Speaker Classifiers}	
	\label{Comparison of style transfer without style/speaker classifier}	
 As mentioned in Section~\ref{Introduction}, our proposed model can be considered as a variant of our previous model~\cite{An2021improving}, which also uses style encoder and speaker encoder but in a different way. Here, we denote the previous model as the Pre-model\textsuperscript{\cite{An2021improving}}. Specially, in the Pre-model\textsuperscript{\cite{An2021improving}}, no style classifier and speaker classifier are used in the corresponding style and speaker encoders, hence no constraints of style and speaker classification losses are imposed in the model training process. The Pre-model\textsuperscript{\cite{An2021improving}} is therefore unable to distinguish the different styles and speakers in a direct manner. In the experiments, we compare the new proposed approach and the Pre-model\textsuperscript{\cite{An2021improving}} with ABX subjective tests of style similarity and speaker similarity. The ABX results are shown in Fig.~\ref{fig:withoutclassifierabx}. We show that listeners give higher preference to the new proposed model which has improved the corresponding style and speaker similarities. Similar to Section~\ref{Style similarity}, we also show the scatter plots of the style embedding space for the Pre-model\textsuperscript{\cite{An2021improving}} and the proposed model. As shown in Fig.~\ref{fig:tsneclassifers}, our proposed model produces more compact and highly separable clusters than the Pre-model\textsuperscript{\cite{An2021improving}} on both seen and unseen style transfer. All evidence shows that the style and speaker classifiers proposed in this paper contribute to better style and speaker similarities for both seen and unseen style transfer, where the different styles and speakers are more distinguishable. The results confirm that the style and speaker classifiers are advantageous to make the transfer process more discriminative in styles and speakers.

\begin{table}[!t]
			\caption{\label{tab:subjectivestyleclassification} {Accuracy of subjective style category classification based on the ablation study of our proposed model.}}
			\centerline{
            %\scalebox{0.75}{
				\begin{tabular}{|c|ccc|ccc|}
					\hline
					\multirow{2}*{Objective} & \multicolumn{3}{c|}{\underline{Seen style transfer}} & \multicolumn{
3}{c|}{\underline{Unseen style transfer}} \\
					   & {R2C} & {R2P}  & {R2G}   & {TR2C} & {TR2P}  & {TR2G}  \\
					\hline
					$\mathcal{L}_{rec}$      &0.64    &0.63  & 0.61   &0.56  &0.54  & 0.53    \\
					+$\mathcal{L}_{adv}$     &0.70    &0.68  & 0.67   &0.63  &0.62  & 0.60    \\			
					+$\mathcal{L}_{dis}$     &0.82    &0.80  & 0.78   &0.74  &0.73  & 0.71    \\
					+$\mathcal{L}_{cyc}$     &0.85    &0.83  & 0.82   &0.79  &0.75  & 0.74    \\
					+$\mathcal{L}_{stycls}$+$\mathcal{L}_{spkcls}$  &\textbf{0.94} &\textbf{0.93} &\textbf{0.92}  &\textbf{0.88} &\textbf{0.87} &\textbf{0.84} \\
					\hline
				\end{tabular}
			}
           % }
		\end{table}

\begin{table}[!t]
			\caption{\label{tab:subjectivespeakerclassification} {Accuracy of subjective speaker category classification based on the ablation study of our proposed model.}}
			\centerline{
            %\scalebox{0.75}{
				\begin{tabular}{|c|ccc|ccc|}
					\hline
					\multirow{2}*{Objective} & \multicolumn{3}{c|}{\underline{Seen style transfer}} & \multicolumn{
3}{c|}{\underline{Unseen style transfer}} \\
					   & {R2C} & {R2P}  & {R2G}   & {TR2C} & {TR2P}  & {TR2G}  \\
					\hline
					$\mathcal{L}_{rec}$      &0.62    &0.60  & 0.59   &0.57  &0.56  & 0.54    \\
					+$\mathcal{L}_{adv}$     &0.69    &0.68  & 0.66   &0.63  &0.61  & 0.60    \\			
					+$\mathcal{L}_{dis}$     &0.76    &0.75  & 0.73   &0.70  &0.69  & 0.67    \\
					+$\mathcal{L}_{cyc}$     &0.86    &0.84  & 0.83   &0.81  &0.80  & 0.78    \\
					+$\mathcal{L}_{stycls}$+$\mathcal{L}_{spkcls}$  &\textbf{0.95} &\textbf{0.93} &\textbf{0.92}  &\textbf{0.90} &\textbf{0.89} &\textbf{0.87} \\
					\hline
				\end{tabular}
			}
           % }
		\end{table}
	
	\subsection{Effect of Different Objectives on Style Transfer}	
	\label{Effect of different objectives on style transfer}
In this paper, we conduct ablation studies on seen and unseen speech style transfer to validate the effectiveness of different objectives. Specifically, we respectively conduct two subjective tests: a style classification test and a speaker classification test, on the same evaluation sets where the same 15 listeners are asked to focus on the style or speaker identity. The style classification accuracy and the speaker classification accuracy are shown in Table~\ref{tab:subjectivestyleclassification} and Table~\ref{tab:subjectivespeakerclassification}, respectively. For each testing sample, we let the listeners select one style from eight style categories (i.e., reading, broadcasting, talking, story-telling, customer-service, poetry, game, and Taiwanese-reading styles), or select one speaker from eight speaker categories (i.e., seven speakers seen in training and a new speaker unseen in training). We can notice that the addition of the adversarial loss ($\mathcal{L}_{adv}$), the style distortion loss ($\mathcal{L}_{dis}$), the cycle consistency loss ($\mathcal{L}_{cyc}$), and the style and speaker classification losses ($\mathcal{L}_{stycls}$ and $\mathcal{L}_{spkcls}$) can bring substantial gain for the style classification accuracy and the speaker classification accuracy. The combination of all losses achieves the best accuracy, which outperforms the Pre-model\textsuperscript{\cite{An2021improving}} model (i.\,e., without the $\mathcal{L}_{stycls}$ and $\mathcal{L}_{spkcls}$) by a large margin. In addition, the listeners find that it is much easier to distinguish the category of each style and each speaker after the addition of $\mathcal{L}_{stycls}$ and $\mathcal{L}_{spkcls}$.

	\section{Conclusion}
	\label{Conclusion}
	We proposed a novel approach to improve performance of seen and unseen speech style transfer on disjoint, multi-style datasets. More precisely, the introduction of IAF technique has significantly improved the variational inference performance to learn a distinctive and expressive style representation, and the introduction of a well-designed speaker encoder has performed a joint training for learning a discriminative speaker representation. A reconstruction term is used to measure the distortions in both source and target reconstructions, and the adversarial loss for ``fooling" a well-trained discriminator. A style distortion loss has made the style representation of a source utterance closer to the target style representation while a cycle consistency loss ensures the transferred utterance to preserve the speaker identity of the source utterance. The addition of a style classifier and a speaker classifier has further enhanced the style and speaker representations, and made the category of each style and each speaker more distinguishable. We have conducted extensive experiments to analyze the IAF structure, the transfer performance without the style and speaker classifiers, and style transfer in different objectives. Experimental results have demonstrated the superior performance of the proposed approach on both seen and unseen style transfer. In the future, we would incorporate our findings to disentangle style and speaker representations, and further improve the speech style transfer performance. We will also apply our seen and unseen style transfer to other TTS applications.

	\section*{Acknowledgment}
	This work was supported by the National Key Research and Development Program of China (No. 2020AAA0108600).
	\ifCLASSOPTIONcaptionsoff
	\newpage
	\fi
	
	%\clearpage
	%\newpage
	\bibliographystyle{IEEEtran}
	\bibliography{TASLP_YYG-R1-v3}
	
\end{document}